\documentclass[journal]{IEEEtran}
\usepackage{algorithm}
\usepackage{algpseudocode}
\usepackage{cite}
\usepackage{soul,tabularx, booktabs}
\usepackage{epsfig,bm,amsmath,amssymb,graphicx,setspace}
\usepackage[numbers]{natbib}
\usepackage{multirow}
\usepackage{ragged2e}
\usepackage{balance}

\begin{document}
   
    \title{Medical Image Segmentation Using Advanced Unet: VMSE-Unet and VM-Unet CBAM+}
    
  \author{Sayandeep Kanrar, Raja Piyush, Qaiser Razi, Debanshi Chakraborty, Vikas Hassija, GSS Chalapathi,~\IEEEmembership{Senior Member,~IEEE}
   \thanks{Sayandeep Kanrar, Raja Piyush, Debanshi Chakraborty, and Vikas Hassija are with the School of Computer Engineering, Kalinga Institute of Industrial Technology (KIIT) Deemed to be University, Bhubaneswar-751024, Odisha, India (e-mail: 22053357@kiit.ac.in, 20051502@kiit.ac.in, 22051508@kiit.ac.in, vikas.hassijafcs@kiit.ac.in).}%
		\thanks{Qaiser Razi and GSS Chalapathi are with the Department of Electrical and Electronics Engineering, BITS-Pilani, Pilani Campus, India 333031 (e-mail: {p20210070, gssc}@pilani.bits-pilani.ac.in).}}%
	
\maketitle  
   \IEEEpeerreviewmaketitle
\begin{abstract}
In this paper, we present the VMSE U-Net and VM-Unet CBAM+ model, two cutting-edge deep learning architectures designed to enhance medical image segmentation. Our approach integrates Squeeze-and-Excitation (SE) and Convolutional Block Attention Module (CBAM) techniques into the traditional VM U-Net framework, significantly improving segmentation accuracy, feature localization, and computational efficiency. Both models show superior performance compared to the baseline VM-Unet across multiple datasets. Notably, VMSE-Unet achieves the highest accuracy, IoU, precision, and recall while maintaining low loss values. It also exhibits exceptional computational efficiency with faster inference times and lower memory usage on both GPU and CPU. Overall, the study suggests that the enhanced architecture VMSE-Unet is a valuable tool for medical image analysis. These findings highlight its potential for real-world clinical applications, emphasizing the importance of further research to optimize accuracy, robustness, and computational efficiency.
\end{abstract}
    
\section{Introduction}
Image segmentation has changed dramatically, from manual approaches requiring much human labor to advanced machine learning and deep learning methods. Early manual segmentation methods were known for their accuracy but were highly labor-intensive and susceptible to human error, rendering them impractical for large-scale applications. The introduction of machine learning (ML) and convolutional neural networks (CNNs) transformed image analysis, automating the segmentation process and significantly improving accuracy \cite{cite42}. The launch of U-Net in 2015 marked a significant leap in this field, offering a robust encoder-decoder architecture tailored for medical image segmentation, and it has since influenced numerous subsequent models \cite{cite61}. To improve its ability to handle challenging segmentation tasks and provide competitive performance for medical imaging applications like tumor identification and organ segmentation, VM-Unet built upon this by including variational approaches \cite{cite43}.

The need for automation and increased accuracy in image segmentation, which became essential in fields like medical imaging, self-driving cars, and industrial inspections, led to the creation of machine learning algorithms, especially CNNs, which are excellent at spotting patterns and features that are difficult for humans to notice \cite{cite45}. These networks, including U-Net and VM-Unet, have demonstrated their efficacy in medical applications, especially where annotated datasets are scarce. U-Net’s encoder-decoder structure facilitated precise localization and segmentation, even with limited data. At the same time, VM-Unet incorporated a Visual State Space (VSS) block to capture long-range dependencies and improve segmentation accuracy in complex medical images \cite{cite49}. These innovations have played a vital role in improving diagnostic tools and treatment planning in healthcare by enabling accurate image analysis.

To enhance the VM-Unet architecture and address its limitations, our study introduces two novel models, VMSE-Unet and VM-Unet CBAM+, which significantly integrate advanced attention mechanisms to improve segmentation accuracy and computational efficiency. Unlike transformer-based architectures that excel at capturing long-range dependencies but suffer from high computational costs, our models achieve a balance between accuracy and efficiency by incorporating lightweight yet powerful mechanisms such as Squeeze-and-Excitation (SE) Attention and Convolutional Block Attention Module (CBAM) \cite{cite48}. SE Attention recalibrates channel-wise feature responses dynamically, enabling the model to prioritize critical features while reducing redundancy. CBAM further enhances segmentation accuracy by introducing spatial and channel attention, allowing the model to focus on significant areas of the image adaptively. These enhancements ensure superior feature extraction and localization compared to existing methods, including multi-scale approaches like MS-UNet and adaptive architectures such as Adaptive Mamba U-Net, which often require extensive hyperparameter tuning or fail to generalize across diverse datasets \cite{cite1}. By evaluating our models on benchmark datasets such as MICCAI 2009, Kvasir-SEG, and BUS, we consistently improve metrics like accuracy, Intersection over Union (IoU), precision, recall, inference time, and memory usage. The proposed architectures outperform the baseline VM-Unet and set a new benchmark for medical image segmentation by delivering robust performance with reduced computational overhead, making them ideal for real-world clinical applications where accuracy and efficiency are paramount \cite{cite6}.
\vspace{0.85cm}
\subsection{Motivation}
 A critical limitation observed across the existing models is their substantial computational demand. While these methods achieve state-of-the-art segmentation performance through  \cite{cite62}, transformer-based architectures \cite{cite63}, and dynamic adaptation layers \cite{cite64}, their reliance on high memory and processing power significantly constrains their applicability. Such computational inefficiency hinders their deployment in real-time scenarios and on resource-constrained platforms, such as edge devices or systems with limited hardware capabilities. This limitation underscores the need for novel solutions and motivates the development of a new model that combines state-of-the-art segmentation performance with significantly reduced computational overhead. The proposed model is designed to achieve an optimal trade-off between segmentation accuracy and efficiency, enabling widespread deployment and accessibility for practical, real-world applications without concerns regarding computational resource constraints.
\subsection{Contribution}
This paper introduces two new models, VMSE-Unet and VM-Unet CBAM+, built on the VM-Unet architecture with SE blocks and CBAM attention mechanisms. Evaluated on MICCAI 2009, Kvasir-SEG, and BUS datasets, the models achieve superior segmentation accuracy, IoU, and recall while reducing inference time and memory usage. VMSE-Unet demonstrates the best IoU and efficiency, making this model ideal for real-time and resource-constrained applications, advancing practical medical image segmentation.
\subsection{Organization}
   The rest of this paper is organized as follows. An introduction to the concept and recent studies on the quantization of VM-Unet are covered in Section II. Background details on the VM-Unet model and its enhancements are given in Section III. The dataset and preprocessing methods used to refine the model are described in depth in Section IV. Section V looks at the training strategy and the configurations used to train our proposed model. The experimental results and evaluation metrics are presented in Section VI. Finally, the paper's conclusion is given in Section VII.
\section{Related Works}
    Medical image segmentation has witnessed significant advancements in integrating deep learning techniques. Among these, the VM-UNet series improves segmentation accuracy and efficiency through innovative approaches. This section reviews several prominent papers that have substantially contributed to advancing VM-UNet for medical image segmentation. Table \ref{table1} outlines the principal contributions of prior research on VM-Unet.

M. Zhang \textit{et al.} \cite{cite10} enhanced the original VM-UNet architecture by introducing the VM-UNetV2 model, which incorporates advanced mechanisms such as the Vision Scanning Selective (VSS) block and the Spatial-Domain Interaction (SDI) module. These components enabled the model to extract more intricate features within complex medical images. Additionally, deeper network layers and improved training strategies allowed VM-UNetV2 to demonstrate superior segmentation performance on datasets like ISIC 2017 and ISIC 2018 with higher metrics such as Mean Intersection over Union (mIoU) and Dice Similarity Coefficient (DSC).

   \begin{table*}[!t]
    \caption{Related Work on Vision Mamba UNet}
    \centering
    \resizebox*{1\textwidth}{!}{
        \begin{tabular}{|l|l|p{8cm}|p{6cm}|}
            \hline
            Year & Reference & Contributions & Limitations \\
            \hline
            2021 & Devidas T. Kushnure \textit{et al.} \cite{cite21} & Developed MS-UNet, integrating multi-scale feature extraction and feature recalibration for improved liver and tumor segmentation in CT images, achieving high accuracy on the 3Dircadb dataset. & Limited generalizability to datasets outside CT imaging due to reliance on specific multi-scale features. \\
            \hline
            2022 & Xiangyi Yan \textit{et al.} \cite{cite20} & Proposed AFTer-UNet, combining Axial Fusion Transformer with UNet for enhanced long-range dependency modeling in medical image segmentation, outperforming existing methods on benchmark datasets. & High computational cost associated with transformer-based architectures limits deployment in resource-constrained environments. \\
            \hline
            2024 & M. Zhang \textit{et al.} \cite{cite10} & VM-UNetV2: Enhanced feature extraction using Vision Scanning Selective (VSS) block and Spatial-Domain Interaction (SDI) module, improving segmentation performance on ISIC datasets. & Performance improvements are dataset-specific and may not generalize well to non-ISIC datasets. \\
            \hline
            2024 & J. Wang \textit{et al.} \cite{cite11} & Large Window-based Mamba UNet: Introduced large window operations to capture long-range dependencies beyond convolution and self-attention, enhancing segmentation accuracy and robustness. & Large window operations increase memory requirements, limiting scalability to high-resolution images. \\
            \hline
            2024 & Z. Wang \textit{et al.} \cite{cite12} & Efficient Mamba U-Net: Optimized computational efficiency with depthwise separable convolutions and efficient block designs, reducing model complexity while maintaining high segmentation performance. & Simplified architecture may compromise segmentation accuracy for highly complex medical images. \\
            \hline
            2024 & C. Yuan \textit{et al.} \cite{cite13} & Adaptive Mamba U-Net: Incorporated dynamic adaptation layers for multi-modal image processing, adjusting parameters based on input modality to improve segmentation across different medical imaging types. & Requires extensive hyperparameter tuning for different modalities, increasing development time. \\
            \hline
            2024 & \textbf{Our Work} & We introduce VMSE-Unet and VM-Unet CBAM+, two advanced deep-learning models designed for medical image segmentation. Building on the traditional VM-UNet, these models incorporate SE blocks and CBAM attention mechanisms to enhance segmentation accuracy and efficiency, ensuring improved performance with reduced computational requirements.& -- \\
            \hline
        \end{tabular}
        }
    \label{table1}
\end{table*}

J. Wang \textit{et al.}  \cite{cite11} developed a large-window approach for the Mamba UNet, which extends beyond traditional convolutional layers and self-attention mechanisms. By incorporating large window operations, this model captures long-range dependencies more effectively, enhancing the ability to delineate anatomical structures and pathological regions. These advancements significantly improved segmentation performance across diverse medical imaging modalities.
Z. Wang \textit{et al.}  \cite{cite12} addressed computational challenges by creating a highly efficient U-Net architecture. Their work focused on optimizing the Mamba U-Net by employing depthwise separable convolutions and lightweight block designs. These enhancements reduced computational overhead and model complexity while maintaining high segmentation performance, making the solution ideal for resource-constrained environments and real-time applications.

C. Yuan \textit{et al.}  \cite{cite13} presented an adaptive approach that introduced dynamic adaptation layers to handle multi-modal medical images effectively. These layers adjust the network's parameters based on input modalities, significantly enhancing the versatility of the Mamba U-Net model. This adaptability made the model capable of processing MRI, CT, and ultrasound images with consistent and high-quality segmentation results, improving its usability in clinical applications.
Yan \textit{et al.}  \cite{cite20} developed AFTer-UNet by integrating Axial Fusion Transformer (AFT) mechanisms with the traditional U-Net. This model enhances the fusion of long-range dependencies and fine-grained representation learning. Superior performance was achieved in segmenting complex medical images, particularly in capturing contextual and detailed image features, as validated with benchmark datasets.
Kushnure \textit{et al.}  proposed MS-UNet \cite{cite14}, a model designed with multi-scale feature extraction and recalibration techniques tailored for liver and tumor segmentation in CT images. By employing a multi-scale approach to capture both global and local features, along with channel-wise feature recalibration, this model achieved exceptional accuracy rates on the 3Dircadb dataset, with Dice scores of 97.13 for liver segmentation and 84.15 for tumor segmentation, outperforming existing methods.

These works significantly contribute to advancements in medical image segmentation by designing novel architectures, optimizing efficiency, and improving adaptability across imaging modalities. The progression from U-Net to Vision Mamba U-Net reflects a sustained effort to refine segmentation techniques that ensure better accuracy and efficiency, surpassing prior limitations and setting new benchmarks in the field. The combination of innovative architectures and sophisticated computational methodologies continues to drive breakthroughs in medical image segmentation \cite{cite14}.

\section{Refinements and Methodology}
In this section, we discussed the VM-Unet model and the enhancements done to it accordingly, along with its relevance.

 \begin{figure*}
    \centering
    \includegraphics[width=150mm]{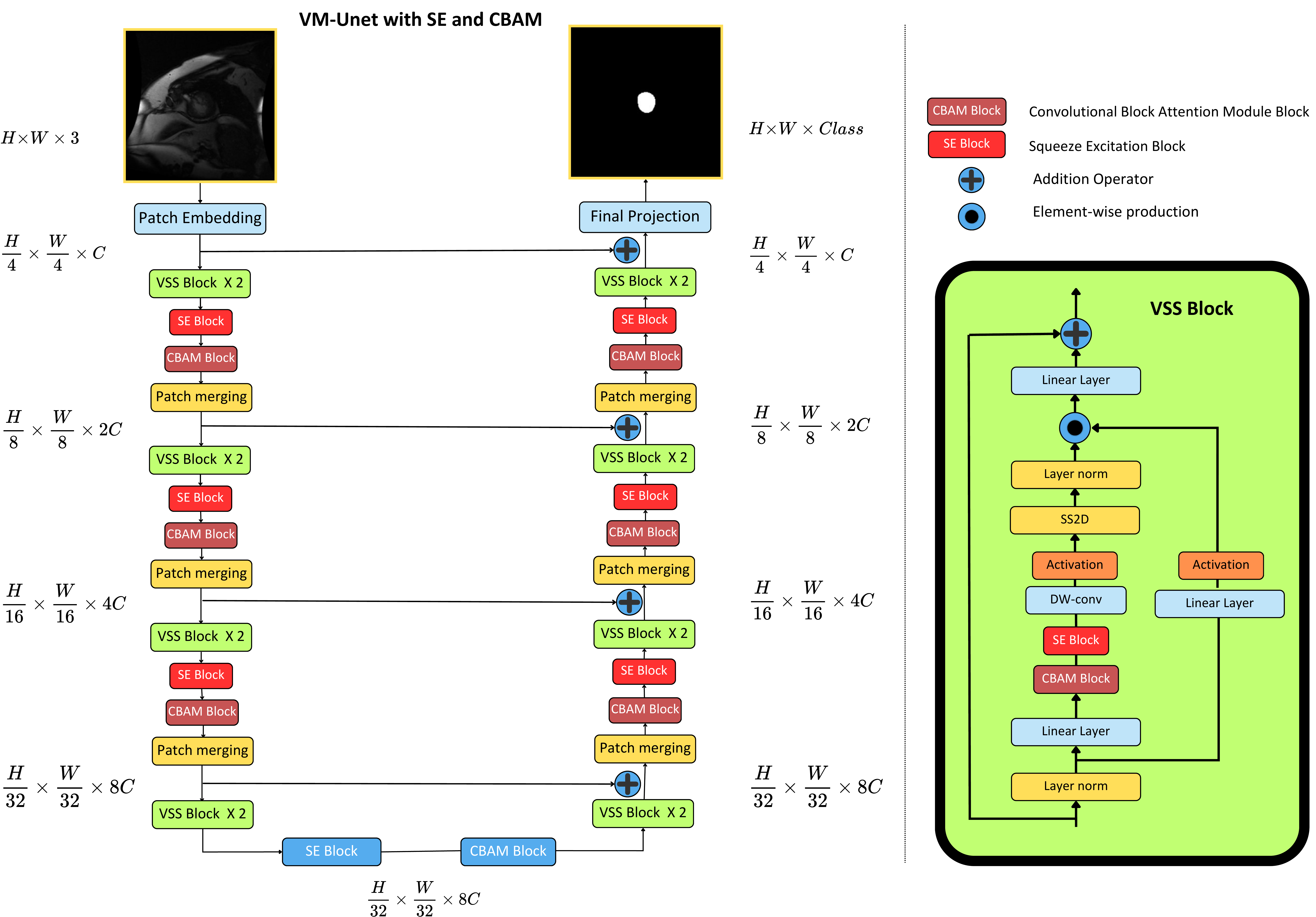}
    \caption{Working diagram of our proposed model.}
    \label{model}
\end{figure*}
 
  \subsection{Model Overview :}
A sophisticated model created especially for medical image segmentation is VM-Unet \cite{cite15}. Building on the fundamental U-Net design, it offers several important improvements that significantly boost its functionality, especially when processing intricate medical pictures. The core architecture of VM-Unet retains the encoder-decoder structure, where the encoder captures essential features through downsampling, and the decoder reconstructs the segmented image via upsampling \cite{cite16}. A key advancement in VM-Unet is the addition of the Visual State Space (VSS) block, which is highly effective in capturing long-range dependencies and contextual details, thereby overcoming the drawbacks of conventional CNNs and transformers. This block models the image as a state space, allowing the network to efficiently capture and utilize global context, which is crucial for accurate segmentation \cite{cite17}.

\subsection{Enhancements in VM-Unet :}
Several enhancements in VM-Unet contribute to its superior
performance in biomedical image segmentation :
\begin{enumerate}
    \item \textit{Visual State Space (VSS) Block : }The VSS block is a pivotal enhancement in VM-Unet, designed to capture long-range dependencies and contextual information within images \cite{cite51}. This block models the image as a state space, allowing the network to capture and utilize global context efficiently. This capability addresses the limitations of traditional CNNs and transformers, which often struggle with long-range interactions, thereby improving the accuracy of segmentation in complex medical images \cite{cite52}.
    \item \textit{Variational Methods : }VM-Unet integrates variational methods to enhance segmentation accuracy by handling uncertainty and variability in medical images. This probabilistic modeling approach is particularly beneficial in medical imaging, where variations in anatomy and pathology can be significant \cite{cite53}. By incorporating these methods, VM-Unet can provide more reliable and precise segmentation results, which are crucial for accurate diagnosis and treatment planning \cite{cite54}.
    \item \textit{SE Attention Mechanism : }By recalibrating channel-wise feature responses, the integration of SE (Squeeze-and-Excitation) Attention methods in VM-Unet aids the network in concentrating on the most crucial features \cite{cite55}. This enhancement allows the model to prioritize important regions within the image, improving segmentation performance, especially in complex and detailed medical images \cite{cite56}. The SE block starts with a global pooling operation to generate channel-wise statistics:
\begin{equation}
z_c = \frac{1}{H \times W} \sum_{i=1}^{H} \sum_{j=1}^{W} F_{c, i, j}
\end{equation}
where \( z_c \) represents the aggregated global context for channel \( c \).

Next, these channel-wise statistics are passed through two fully connected layers with a non-linear activation function to produce recalibration weights:
\begin{equation}
s_c = \sigma(W_2 \cdot \text{ReLU}(W_1 \cdot z_c))
\end{equation}
where \( W_1 \) and \( W_2 \) are learnable weight matrices, and \( \sigma \) denotes the sigmoid function.

The recalibrated weights are applied to the original feature map via element-wise multiplication:
\begin{equation}
\hat{F}_c = F_c \cdot s_c
\end{equation}
where \( \hat{F}_c \) is the channel-refined feature map.
    \item \textit{Convolutional Block Attention Module (CBAM) : }Through the integration of both spatial and channel attention, this enhances the attention mechanism even more, \cite{cite57}. By increasing the model's sensitivity to significant areas and characteristics in the image, this dual attention technique raises the segmentation's overall accuracy and resilience \cite{cite58}. Adding CBAM to VM-Unet is beneficial since it can dynamically change focus according to the properties of the input image. Channel attention is computed by aggregating spatial information using average pooling and max pooling:
\begin{equation}
z_{\text{avg}, c} = \frac{1}{H \times W} \sum_{i=1}^{H} \sum_{j=1}^{W} F_{c, i, j}, \quad
z_{\text{max}, c} = \max_{i, j} F_{c, i, j}
\end{equation}
where \( z_{\text{avg}, c} \) is the average-pooled value for channel \( c \), summarizing spatial features. \( z_{\text{max}, c} \) is the max-pooled value for channel \( c \), capturing the most prominent spatial feature.
 \( F_{c, i, j} \) represents the input feature map value for channel \( c \), at spatial location \( (i, j) \). \( H, W \) are the height and width of the input feature map.

The pooled outputs are passed through fully connected layers to compute channel-wise attention weights:
\begin{equation}
s_c = \sigma(W_2 \cdot \text{ReLU}(W_1 \cdot z_{\text{avg}}) + W_2 \cdot \text{ReLU}(W_1 \cdot z_{\text{max}}))
\end{equation}
where \( s_c \) is the channel attention weight for channel \( c \), scaled to \([0, 1]\). \( z_{\text{avg}}, z_{\text{max}} \) are the aggregated spatial features from the previous step. \( W_1, W_2 \) are the weight matrices for the two fully connected layers. \( \text{ReLU} \) is the Rectified Linear Unit activation function and \( \sigma \) is the Sigmoid activation function.

These weights are used to refine the feature map via element-wise multiplication:
\begin{equation}
\hat{F}_{\text{channel}, c, i, j} = F_{c, i, j} \cdot s_c
\end{equation}
where \( \hat{F}_{\text{channel}, c, i, j} \) represents the refined feature map value for channel \( c \), at location \( (i, j) \).  \( F_{c, i, j} \) is the original feature map value for channel \( c \), at location \( (i, j) \). \( s_c \) is the channel attention weight for channel \( c \).

Spatial attention aggregates channel information using average pooling and max pooling:
\begin{equation}
F_{\text{avg}}^{\text{spatial}} = \frac{1}{C} \sum_{c=1}^{C} F_{c, i, j}, \quad
F_{\text{max}}^{\text{spatial}} = \max_{c} F_{c, i, j}
\end{equation}
where \( F_{\text{avg}}^{\text{spatial}} \) represents average-pooled spatial feature map across all channels. \( F_{\text{max}}^{\text{spatial}} \) is the max-pooled spatial feature map across all channels. \( F_{c, i, j} \) is the input feature map value for channel \( c \), at location \( (i, j) \). \( C \) is the total number of channels in the feature map.
\begin{figure*}
    \centering
    \includegraphics[width=145mm]{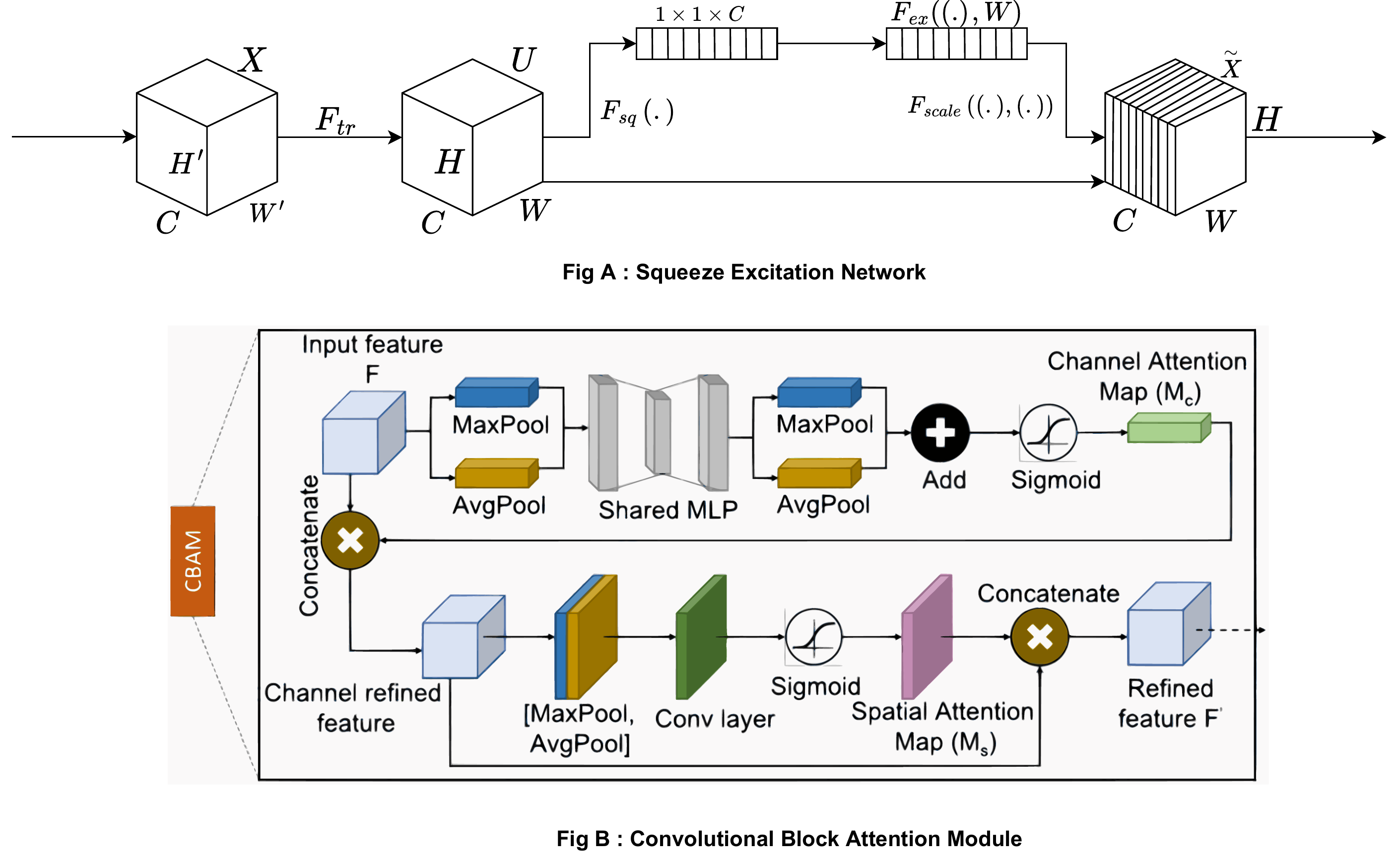}
    \caption{Diagram of Squeeze Excitation Block and Convolutional Block Attention Module. }
    \label{secbam}
\end{figure*}
\item \textit{Proposed Model Architecture  : }In this study, we propose an enhanced version of the Vision Mamba U-Net (VM-UNet) architecture, which integrates advanced attention mechanisms, namely the Squeeze-and-Excitation (SE) block and the Convolutional Block Attention Module (CBAM) as shown in Fig. \ref{model}. These modifications aim to improve feature extraction and segmentation accuracy in complex medical image datasets.

The baseline VM-UNet architecture follows a standard encoder-decoder structure with skip connections. The feature map \( F \) at any layer is computed as:
\begin{equation}
F = \text{Conv}(X) + \text{UpConv}(F_{\text{skip}})
\end{equation}

where \( X \) is the input feature map and \( \text{Conv}(\cdot) \) is the convolution operation applied to the input. \( \text{UpConv}(\cdot) \) is the transposed convolution for upsampling and \( F_{\text{skip}} \) represents the feature map from the encoder passed via skip connections.

The SE block enhances the feature map by recalibrating channel-wise responses, as explained in Fig. \ref{secbam}. After applying SE, the feature map \( F_{\text{SE}} \) is computed as:
\begin{equation}
F_{\text{SE}, c} = F_c \cdot \sigma(W_2 \cdot \text{ReLU}(W_1 \cdot z_c))
\end{equation}
where \( F_c \) represents the original feature map for channel \( c \), and \( z_c \) is the channel-wise global context obtained through global average pooling, defined as
\begin{equation}
z_c = \frac{1}{H \times W} \sum_{i=1}^{H} \sum_{j=1}^{W} F_{c, i, j}
\end{equation}
where \( W_1 \) and \( W_2 \) are learnable weight matrices for the fully connected layers, \( \text{ReLU} \) is the Rectified Linear Unit activation function, and \( \sigma \) denotes the sigmoid function, which scales the attention weights to the range \([0, 1]\).

CBAM applies both channel and spatial attention sequentially, as shown in Fig. \ref{secbam}.  The output feature map \( F_{\text{CBAM}} \) is computed as:
\begin{equation}
F_{\text{CBAM}} = F \cdot \sigma(\text{Conv}([F_{\text{avg}}^{\text{spatial}}, F_{\text{max}}^{\text{spatial}}]))
\end{equation}
where \( F \) is the input feature map, and \( [F_{\text{avg}}^{\text{spatial}}, F_{\text{max}}^{\text{spatial}}] \) represents the concatenated feature maps obtained from average pooling and max pooling across all channels. The average-pooled spatial features are computed as 
\begin{equation}
F_{\text{avg}}^{\text{spatial}} = \frac{1}{C} \sum_{c=1}^{C} F_{c, i, j}
\end{equation}
and the max-pooled spatial features are computed as
\begin{equation}
 F_{\text{max}}^{\text{spatial}} = \max_{c} F_{c, i, j}
\end{equation}
The convolution operation \( \text{Conv}(\cdot) \) is applied to the concatenated features, and \( \sigma \) denotes the sigmoid activation function, which produces spatial attention weights.
 Experimental evaluation will validate the effectiveness of these modifications in addressing complex segmentation tasks, advancing the development of accurate and efficient medical image analysis frameworks.
\end{enumerate}

\section{Dataset and Preprocessing}
This section explains the preprocessing steps taken to prepare the data for training and the dataset used to fine-tune the Vm-Unet model.
\vspace{-1.0em}
\subsection{Dataset Description}

The MICCAI 2009 dataset \cite{cite67}, a medical imaging dataset made available for the MICCAI (Medical Image Computing and Computer Assisted Intervention) 2009 workshop, was the primary data source we used for our investigation \cite{cite31}. This dataset contains multimodal brain pictures with an emphasis on brain tumor segmentation and was created especially for study in the field of medical image processing. Additionally, for comparison purposes, we have considered the Kvasir-SEG dataset \cite{cite65} and the BUS synthetic dataset (Breast Ultrasound synthetic pictures) \cite{cite66}.
\begin{figure}
    \centering
    \includegraphics[width=85mm]{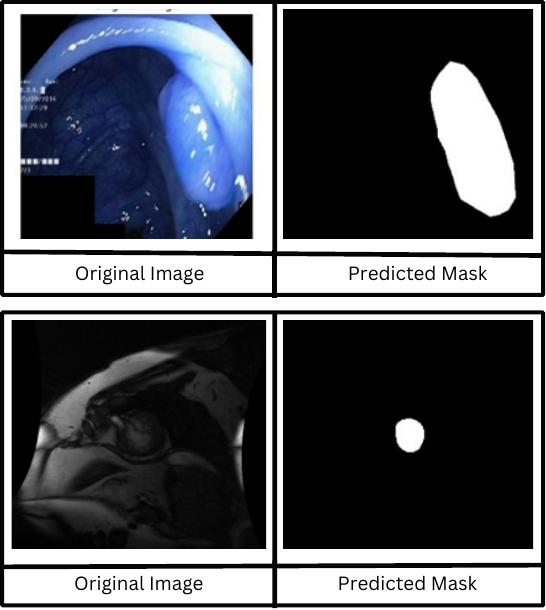}
    \caption{Some perfect predicted masks from VMSE Unet.}
    \label{mask}
\end{figure}
\vspace{-1.0em}
\subsection{Data Collection}
The MICCAI 2009 dataset is sourced from the National Institute of Health 
library and has undergone some custom modifications and pre-processing, ensuring ease of access and compatibility with the tools and frameworks we have used for model training. The Kvasir-SEG and BUS synthetic datasets have been sourced from Kaggle.
\vspace{-0.8em}
\subsection{Data Preprocessing}
The training and testing sets of the raw MRI data, along with the matching ground truth annotations, were obtained in DICOM format.

    Inputs were standardized to zero mean and unit standard deviation, reducing intensity variations and accelerating network convergence. The following is how the normalization is used:
   
\begin{equation}
x' = \frac{x - \mu}{\sigma}
\end{equation}
   where \( x \) represents the original image pixel values and \( x' \) represents normalized pixel values.\( \mu \) is the mean of the pixel values and \( \sigma \) is the standard deviation of the pixel values.
   
Input images were resized to \(256 \times 256\) pixels for network standardization, with the region of interest-focused cropping to preserve anatomical details despite cardiac image size heterogeneity. This can be shown as:
  \begin{equation}
x_{\text{resized}} = \text{resize}(x, 256, 256)
\end{equation}
    where \( x \) is the original image and \( x_{\text{resized}} \) is the resized image with dimensions \( 256 \times 256 \).

    The cropping operation is focused on a region of interest (ROI):
    \begin{equation}  
    x_{\text{cropped}} = x_{\text{resized}}[x_{\text{ROI}}, y_{\text{ROI}}]
    \end{equation}
    where \( x_{\text{ROI}} \) and \( y_{\text{ROI}} \) coordinates for the region of interest in the resized image.

  Several data augmentation methods were used to improve model robustness and reduce overfitting. These included scaling, flipping, random rotations, and elastic deformations. For instance:

         Rotation by angle \( \theta \):

        \begin{equation}  
        x_{\text{rotated}} = \text{rotate}(x, \theta)
        \end{equation}

         Scaling by factor \( s \):

        \begin{equation}  
        x_{\text{scaled}} = \text{scale}(x, s)
        \end{equation}

         Elastic Deformation using a displacement field \( \delta(x) \):

        \begin{equation}  
        x_{\text{deformed}} = x + \delta(x)
        \end{equation}

    A smoothing or denoising filter \( F \), such as Gaussian filtering, is applied as follows:
    \begin{equation} 
    x_{\text{filtered}} = F(x)
    \end{equation}
    where \( F \) represents a filter function (e.g., Gaussian) applied to reduce noise.

The ground truth segmentation masks were binarized to delineate the heart and associated structures from the background. This binarization was crucial for the supervised learning approach employed by VM-Unet variants. The binarization process is given by:
    \begin{equation} 
    m' = 
    \begin{cases} 
      1 & \text{if } m \geq T \\
      0 & \text{if } m < T 
    \end{cases}
   \end{equation} 
    where \( m \) are the original mask pixel values and \( m' \) represent binarized mask values, with 1 indicating the segmented region and 0 indicating the background. \( T \) is the threshold value for binarization.

Our models were trained on modified MICCAI 2009 (3,000 cardiac MRI images with masks), Kvasir-SEG (1,000 polyp images with masks), and BUS (780 categorized breast ultrasound images) datasets, each partitioned into $70:15:15$ training, validation, and testing subsets. Some exemplary mask generations from VMSE-UNet are shown in Fig. \ref{mask}.

\section{Methodology and Training}

This section details procedures to optimize our improved VM-Unet model's performance for medical image segmentation, including hyperparameter tuning (learning rate, batch size), computational resource allocation, and accuracy-enhancing techniques like boundary-aware augmentation.

\vspace{-1.0em}

\subsection{Training Setup}

\subsubsection{Environment and Tools}

The models were trained on Google Colab and Kaggle, leveraging the Nvidia Tesla A100 GPU and Tesla P100 GPU to expedite the training process. All scripts were implemented using Python with TensorFlow and Keras libraries.

\subsubsection{Training Procedure}

   The training and validation datasets were loaded by generating a list of image IDs from specified directories. Custom data generators were created to manage batch processing efficiently. Three callbacks were established to enhance training efficiency:
    \begin{itemize}
        \item \textit{ModelCheckpoint}: The model with the lowest validation loss was saved.

        \item \textit{LearningRateScheduler}: Adjusted the learning rate dynamically according to a predefined schedule.
        \item \textit{ReduceLROnPlateau}: Reduced learning rate when the validation loss plateaued, improving convergence.
    \end{itemize}
    
   The training was executed with specified epochs. The progress was tracked with the validation data, and the final model was saved upon completion.

\begin{algorithm}[H]
\caption{Train Model}
\textbf{Input:} Model \( M \), Training data \( D_{\text{train}} \), Validation data \( D_{\text{val}} \), Epochs \( E \).\\
\textbf{Output:} Trained model \( M_{\text{final}} \).

\begin{algorithmic}[1]
\Function{TrainModel}{}
    \State Initialize callbacks: ModelCheckpoint, LearningRateScheduler, ReduceLROnPlateau.
    \For{epoch \( e = 1 \) to \( E \)}
        \State Train \( M \) on \( D_{\text{train}} \).
        \State Validate \( M \) on \( D_{\text{val}} \).
        \State Update checkpoints and learning rate.
    \EndFor
    \State Save \( M \) as \( M_{\text{final}} \).
\EndFunction
\end{algorithmic}
\end{algorithm}
\vspace{-1.5em}
\begin{algorithm}[H]
\caption{Evaluate Model}
\textbf{Input:} Trained model \( M_{\text{final}} \), Test data \( D_{\text{test}} \).\\
\textbf{Output:} Metrics \( M_{\text{metrics}} \).

\begin{algorithmic}[1]
\Function{EvaluateModel}{}
    \State Initialize \( M_{\text{metrics}} \).
    \For{each batch \( B \) in \( D_{\text{test}} \)}
        \State Predict outputs for \( B \) using \( M_{\text{final}} \).
        \State Compute and accumulate metrics.
    \EndFor
    \State \Return \( M_{\text{metrics}} \).
\EndFunction
\end{algorithmic}
\end{algorithm}
\begin{table*}[h!]
\centering
\caption{Comparison of Models with different metrics}
\label{tab:vm_unet_metrics}
\begin{tabular}{|l|c|c|c|c|c|c|}
\hline
\textbf{Dataset} & \textbf{Methods} & \textbf{Acc} & \textbf{IoU} & \textbf{Precision} & \textbf{Loss} & \textbf{Recall} \\ \hline
\multirow{3}{*}{\textbf{MICCAI 2009}} 
& VM-Unet & 0.9912 & 0.5577 & 0.7324 & 0.0591 & 0.4633 \\ \cline{2-7}
& \textbf{VM-Unet CBAM+} & \textbf{0.9914} & \textbf{0.5679} & \textbf{0.7395} & \textbf{0.3591} & \textbf{0.4732} \\ \cline{2-7}
& \textbf{VMSE-Unet} & \textbf{0.9919} & \textbf{0.5660} & \textbf{0.7962} & \textbf{0.3616} & \textbf{0.4944} \\ \hline
 
\multirow{3}{*}{\textbf{KVASIR-SEG}}
& VM-Unet & 0.8401 & 0.4361 & 0.8500 & 0.3386 & 0.4049 \\ \cline{2-7}
& \textbf{VM-Unet CBAM+} & \textbf{0.9795} & \textbf{0.9597} & \textbf{0.9650} & \textbf{0.0293} & \textbf{0.9274} \\ \cline{2-7}
& \textbf{VMSE-Unet} & \textbf{0.9678} & \textbf{0.9771} & \textbf{0.9798} & \textbf{0.0628} & \textbf{0.9578} \\ \hline

\multirow{3}{*}{\textbf{BUS Dataset}}
& VM-Unet & 0.9556 & 0.8408 & 0.9165 & 0.2434 & 0.7810 \\ \cline{2-7}
& \textbf{VM-Unet CBAM+} & \textbf{0.9895} & \textbf{0.9648} & \textbf{0.9634} & \textbf{0.0323} & \textbf{0.9307} \\ \cline{2-7}
& \textbf{VMSE-Unet} & \textbf{0.9916} & \textbf{0.9726} & \textbf{0.9803} & \textbf{0.0262} & \textbf{0.9540} \\ \hline
\end{tabular}
\end{table*}

\subsection{Evaluation Criteria}

The model’s performance was assessed using several standard metrics in biomedical image segmentation. This measurement assesses the similarity between the predicted segmentation masks and the actual ones. It is defined as:
\begin{equation}
\text{IoU} = \frac{|Y \cap \hat{Y}|}{|Y \cup \hat{Y}|}
\end{equation}
where \( Y \) represents the ground truth segmentation mask, and \( \hat{Y} \) denotes the predicted segmentation mask.

   The Dice Coefficient, defined as follows, emphasizes the significance of striking a balance between recall and precision in segmentation, much like the Intersection over Union (IoU) does.
     \begin{equation}
    \text{Dice Coefficient} = \frac{2 |Y \cap \hat{Y}|}{|Y| + |\hat{Y}|}
     \end{equation}
    
   While recall evaluates the model's ability to identify all relevant instances in the dataset, precision examines how correctly positive predictions are made. They are characterized as:
    \begin{equation}
    \text{Precision} = \frac{TP}{TP + FP}, \quad \text{Recall} = \frac{TP}{TP + FN}
    \end{equation}
 where \( FP \), \( FN \), and \( TP \) stand for false positives, false negatives, and true positives, respectively.  

Validation loss was tracked during training to evaluate model generalization. Binary Cross-Entropy and Dice Loss were combined in the loss function to quantify segmentation overlap and pixel accuracy, while computational efficiency metrics (inference time and memory usage) on GPU and CPU served as additional evaluation criteria \cite{cite41}.

\section{Experimental Results}

\begin{figure}
    \centering
    \includegraphics[width=98mm]{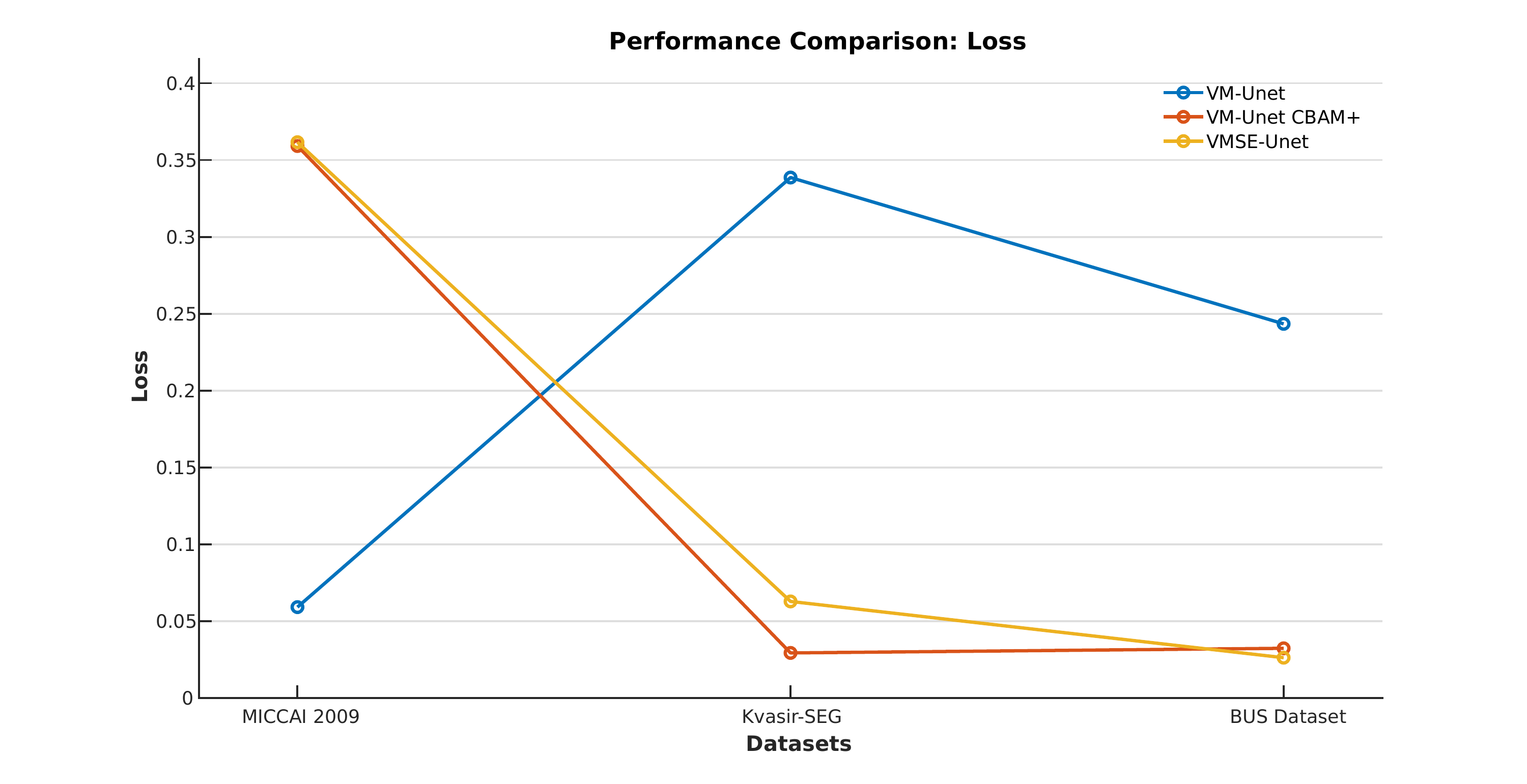}
    \caption{Loss Comparison.}
    \label{loss}
\end{figure}

\begin{figure}
    \centering
    \includegraphics[width=98mm]{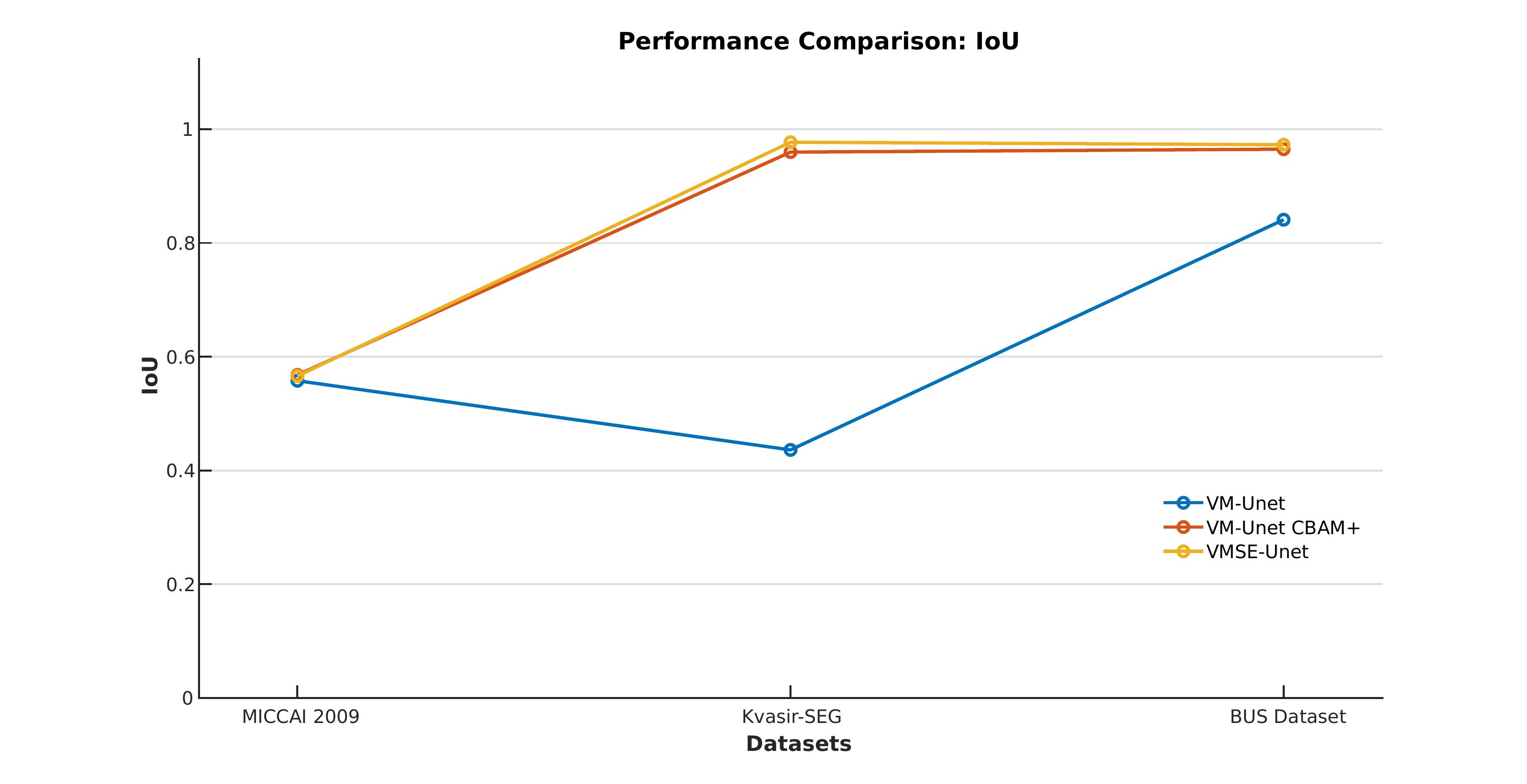}
    \caption{ Intersection Over Union Comparison.}
    \label{iou}
\end{figure}

\begin{figure}
    \centering
    \includegraphics[width=98mm]{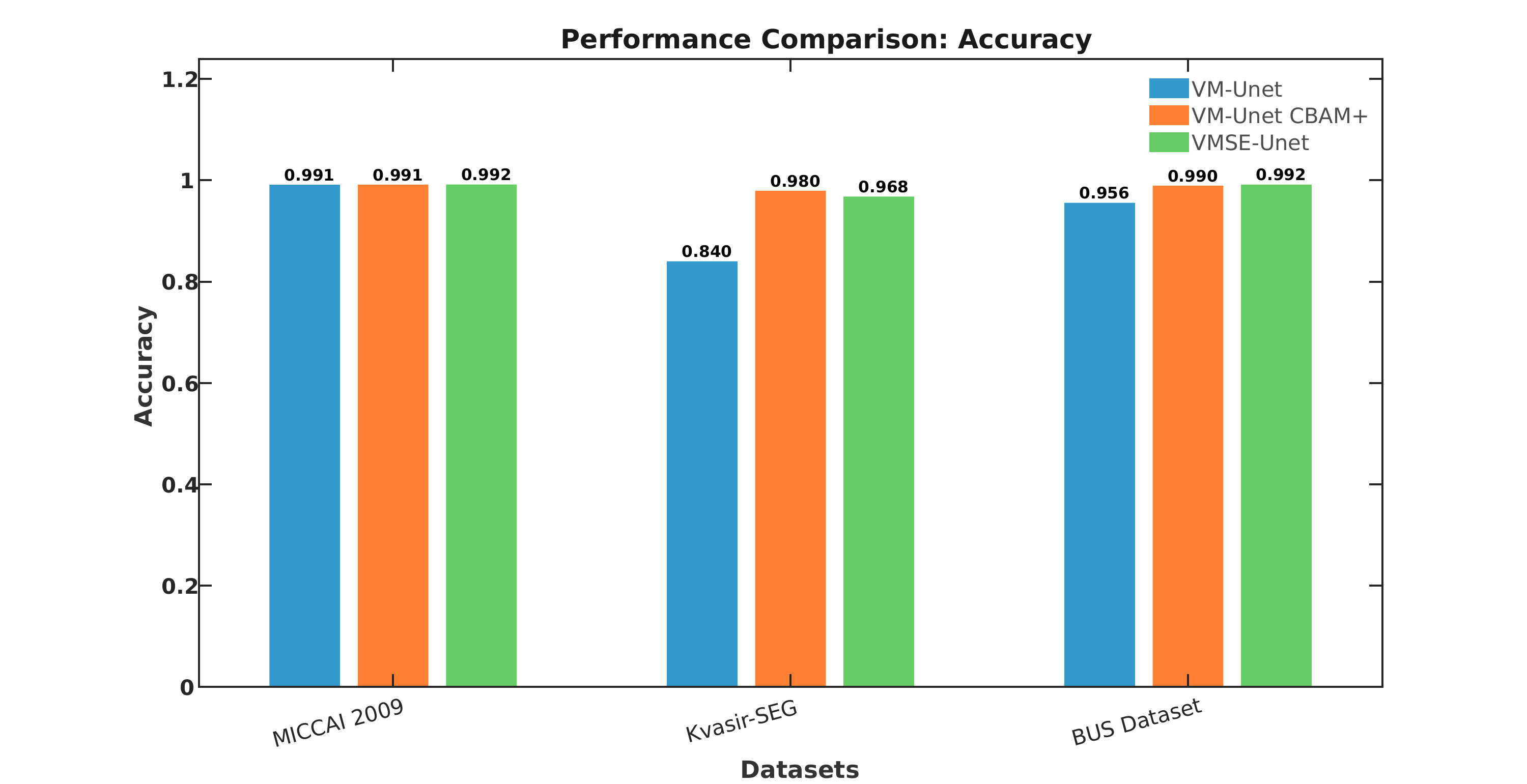}
    \caption{Accuracy Comparison.}
    \label{acc}
\end{figure}

\begin{figure}
    \centering
    \includegraphics[width=98mm]{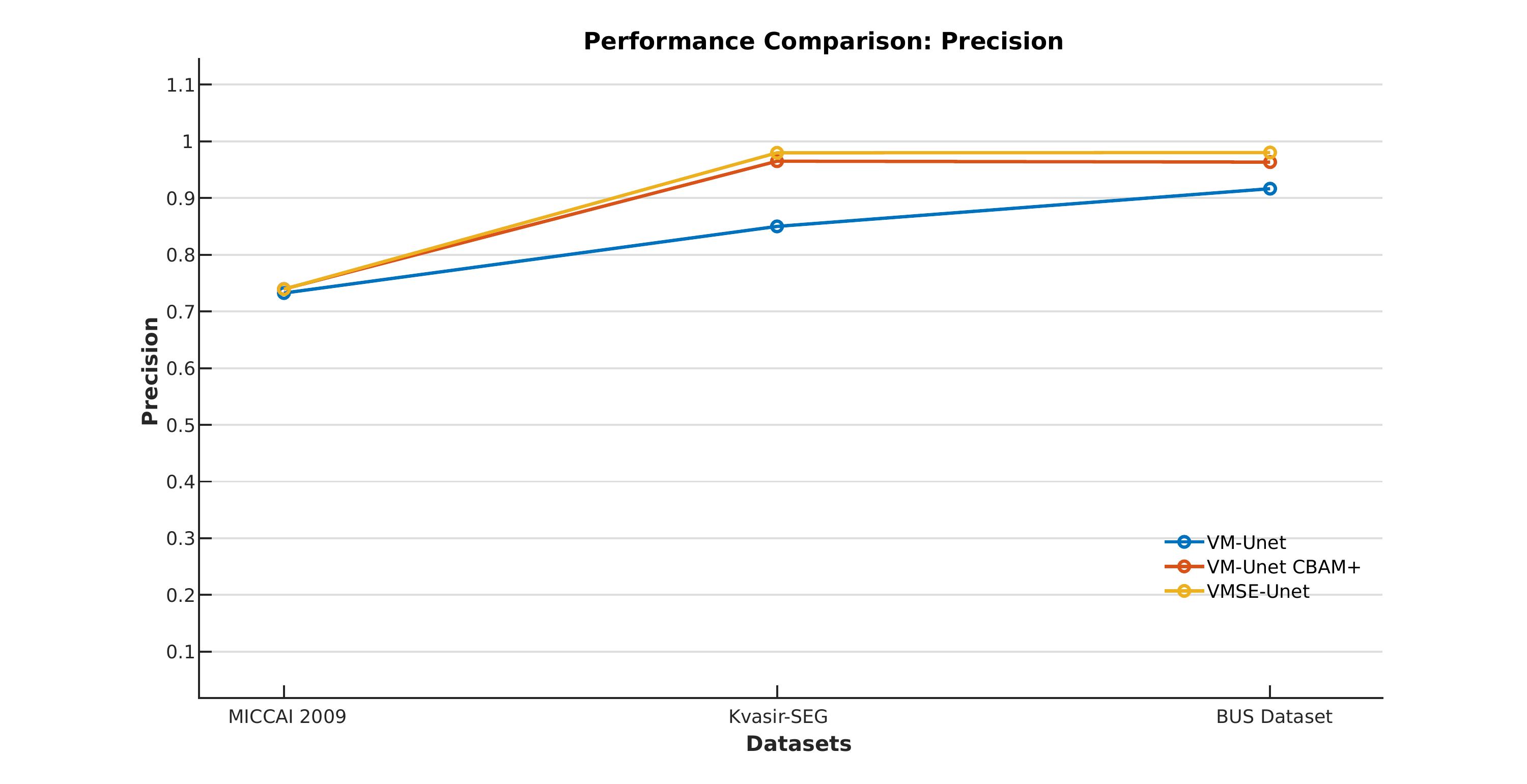}
    \caption{Precision Comparison.}
    \label{precs}
\end{figure}

\begin{figure}
    \centering
    \includegraphics[width=96mm]{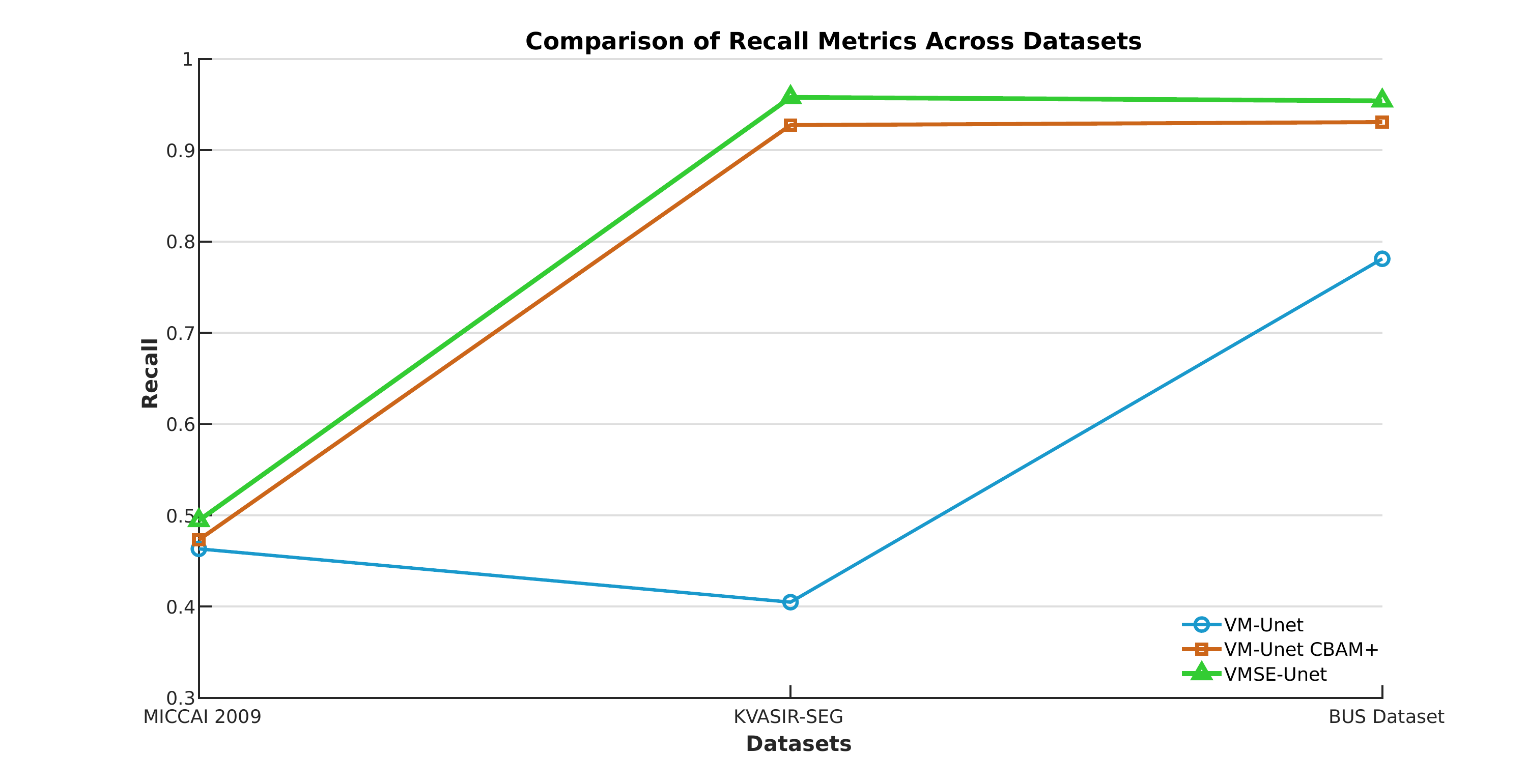}
    \caption{Recall Evaluation.}
    \label{recall}
\end{figure}

The experimental results demonstrate comprehensive performance evaluations across multiple metrics, as illustrated in Figs.~\ref{loss}--\ref{recall}. The comparative analysis reveals several key findings as detailed in Table~\ref{tab:vm_unet_metrics} and Table \ref{tab:inf}.

The Loss Comparison indicates a significant reduction in loss values across datasets, particularly for VM-Unet CBAM+ and VMSE-Unet, demonstrating enhanced model stability as shown in Fig. \ref{loss}. The Intersection over Union (IoU) evaluation shows substantial improvement in segmentation accuracy, with VMSE-Unet achieving superior performance across all datasets, as shown in Fig. \ref{iou}.
The Accuracy Comparison reveals consistent performance improvements, with VMSE-Unet achieving optimal results across all three datasets, as shown in Fig. \ref{acc}. Similarly, the Precision Comparison Fig. \ref{precs}) demonstrates an upward trend in precision metrics, particularly notable in the KVASIR-SEG and BUS datasets.
The Recall Evaluation exhibits a marked improvement in recall metrics for both enhanced architectures. VM-Unet CBAM+ and VMSE-Unet demonstrate substantially higher recall values than baseline VM-Unet, particularly in the KVASIR-SEG and BUS datasets, as shown in Fig. \ref{recall}.

The computational performance analysis, detailed in Table \ref{tab:inf}, demonstrates the superior efficiency of VMSE-Unet across multiple hardware configurations. The architecture achieves remarkable inference speeds, executing predictions in 0.04212 seconds on GPU infrastructure and 1.11716 seconds in CPU environments. Furthermore, VMSE-Unet exhibits exceptional memory optimization, requiring only 2.01 GB and 2.13 GB of memory allocation for GPU and CPU implementations, respectively.
The empirical evidence substantiates VMSE-Unet's position as a state-of-the-art architecture that successfully balances computational efficiency with performance metrics. This optimal resource utilization, coupled with superior segmentation capabilities, positions VMSE-Unet as an ideal candidate for deployment in clinical settings where both computational constraints and diagnostic accuracy are paramount considerations. The architecture's demonstrated ability to maintain high performance while minimizing computational overhead represents a significant advancement in medical image segmentation technology.

\centering
\begin{table*}[h!]
\centering
\caption{Comparison of Models on Inference Time and Memory Usage}
\label{tab:inf}
\begin{tabular}{|l|c|c|c|}
\hline
\textbf{Metric} & \textbf{VM-Unet} & \textbf{VM-Unet CBAM+} & \textbf{VMSE-Unet} \\
\hline
GPU Inference Time (s) & 0.06909 & \textbf{0.06181} & \textbf{0.04212} \\
CPU Inference Time (s) & 1.46353 & \textbf{1.38972} & \textbf{1.11716} \\
Memory Usage (GB)      & 2.59 (GPU), 2.54 (CPU) & \textbf{2.77 (GPU), 2.61 (CPU)} & \textbf{2.01 (GPU), 2.13 (CPU)} \\
\hline
\end{tabular}
\end{table*}

\section{Conclusion}
\justifying
 This study provides a comprehensive evaluation of our proposed models, VM-Unet CBAM+ and VMSE-Unet, for medical image segmentation. Experimental results demonstrate that both models surpass the baseline VM-Unet across multiple datasets and performance metrics. In particular, VMSE-Unet consistently achieves superior performance, exhibiting the highest accuracy, Intersection over Union (IoU), precision, and recall while maintaining minimal loss values. Additionally, VMSE-Unet demonstrates exceptional computational efficiency, characterized by the fastest inference times and the lowest memory consumption on both GPU and CPU. These advancements in performance and efficiency underscore the effectiveness of our proposed enhancements, establishing VMSE-Unet as the optimal model for medical image segmentation tasks. The findings further highlight the potential of VMSE-Unet for real-world clinical applications, where accuracy and computational efficiency are paramount. Future research directions may explore the integration of advanced transformer-based architectures or hybrid attention mechanisms to further refine segmentation performance and efficiency, as well as extend these models for multimodal medical imaging and 3D segmentation applications.

\vspace{1 cm}
\textbf{Author Contributions} Sayandeep and Raja Piyush: Conducted the experiment and authored the main content of the manuscript. Qaiser and Debanshi:  Assisted in preparing the whole manuscript.
Vikas, and GSS: Provided guidance, performed proofreading, and curated the content for the entire paper.
\vspace{0.5 cm}

\textbf{Data Availability} No datasets were generated or analyzed during the
current study.

\vspace{0.5 cm}
\textbf{Funding} This submission was carried out without any external funding sources. The authors declare that they have no financial or nonfinancial interests related to this work.

\vspace{0.5 cm}
\textbf{Declarations}

\vspace{0.5 cm}
\textbf{Ethical Approval}
This article contains no studies with human participants or animals performed by any authors.
\vspace{0.5 cm}

\textbf{Conflict of Interest} The authors declare no competing interests.
\balance
\bibliography{bibliography}

\begin{thebibliography}{10}
\providecommand{\url}[1]{#1}
\csname url@samestyle\endcsname
\providecommand{\newblock}{\relax}
\providecommand{\bibinfo}[2]{#2}
\providecommand{\BIBentrySTDinterwordspacing}{\spaceskip=0pt\relax}
\providecommand{\BIBentryALTinterwordstretchfactor}{4}
\providecommand{\BIBentryALTinterwordspacing}{\spaceskip=\fontdimen2\font plus
\BIBentryALTinterwordstretchfactor\fontdimen3\font minus \fontdimen4\font\relax}
\providecommand{\BIBforeignlanguage}[2]{{%
\expandafter\ifx\csname l@#1\endcsname\relax
\typeout{** WARNING: IEEEtran.bst: No hyphenation pattern has been}%
\typeout{** loaded for the language `#1'. Using the pattern for}%
\typeout{** the default language instead.}%
\else
\language=\csname l@#1\endcsname
\fi
#2}}
\providecommand{\BIBdecl}{\relax}
\BIBdecl

\bibitem{cite42}
A.~Singha, R.~S. Thakur, and T.~Patel, ``Deep learning applications in medical image analysis,'' \emph{Biomedical Data Mining for Information Retrieval: Methodologies, Techniques and Applications}, pp. 293--350, 2021.

\bibitem{cite61}
O.~Ronneberger, P.~Fischer, and T.~Brox, ``U-net: Convolutional networks for biomedical image segmentation,'' in \emph{Medical image computing and computer-assisted intervention--MICCAI 2015: 18th international conference, Munich, Germany, October 5-9, 2015, proceedings, part III 18}.\hskip 1em plus 0.5em minus 0.4em\relax Springer, 2015, pp. 234--241.

\bibitem{cite43}
Z.~Ju and W.~Zhou, ``Vm-ddpm: Vision mamba diffusion for medical image synthesis,'' \emph{arXiv preprint arXiv:2405.05667}, 2024.

\bibitem{cite45}
M.~Sonka, V.~Hlavac, and R.~Boyle, \emph{Image processing, analysis and machine vision}.\hskip 1em plus 0.5em minus 0.4em\relax Springer, 2013.

\bibitem{cite49}
C.~Zhang, A.~Achuthan, and G.~M.~S. Himel, ``State-of-the-art and challenges in pancreatic ct segmentation: A systematic review of u-net and its variants,'' \emph{IEEE Access}, 2024.

\bibitem{cite48}
M.~Oquab, L.~Bottou, I.~Laptev, and J.~Sivic, ``Learning and transferring mid-level image representations using convolutional neural networks,'' in \emph{Proceedings of the IEEE conference on computer vision and pattern recognition}, 2014, pp. 1717--1724.

\bibitem{cite1}
D.~Hein, A.~Bozorgpour, D.~Merhof, and G.~Wang, ``Physics-inspired generative models in medical imaging: A review,'' \emph{arXiv preprint arXiv:2407.10856}, 2024.

\bibitem{cite6}
H.~Zhang, Y.~Zhu, D.~Wang, L.~Zhang, T.~Chen, Z.~Wang, and Z.~Ye, ``A survey on visual mamba,'' \emph{Applied Sciences}, vol.~14, no.~13, p. 5683, 2024.

\bibitem{cite62}
P.~Wu, Z.~Wang, B.~Zheng, H.~Li, F.~E. Alsaadi, and N.~Zeng, ``Aggn: Attention-based glioma grading network with multi-scale feature extraction and multi-modal information fusion,'' \emph{Computers in biology and medicine}, vol. 152, p. 106457, 2023.

\bibitem{cite63}
H.~Xiao, L.~Li, Q.~Liu, X.~Zhu, and Q.~Zhang, ``Transformers in medical image segmentation: A review,'' \emph{Biomedical Signal Processing and Control}, vol.~84, p. 104791, 2023.

\bibitem{cite64}
T.~L{\"u}ddecke and A.~Ecker, ``Image segmentation using text and image prompts,'' in \emph{Proceedings of the IEEE/CVF conference on computer vision and pattern recognition}, 2022, pp. 7086--7096.

\bibitem{cite10}
X.~Xu, X.~Li, and K.~Chen, ``Vm-unetv2: Rethinking vision mamba unet for medical image segmentation,'' \emph{arXiv preprint arXiv:2403.09157}, 2023.

\bibitem{cite21}
D.~T. Kushnure and S.~N. Talbar, ``Ms-unet: A multi-scale unet with feature recalibration approach for automatic liver and tumor segmentation in ct images,'' \emph{Computerized Medical Imaging and Graphics}, vol.~89, p. 101885, 2021.

\bibitem{cite20}
X.~Yan, H.~Tang, S.~Sun, H.~Ma, D.~Kong, and X.~Xie, ``After-unet: Axial fusion transformer unet for medical image segmentation,'' in \emph{Proceedings of the IEEE/CVF winter conference on applications of computer vision}, 2022, pp. 3971--3981.

\bibitem{cite11}
X.~Li, K.~Chen, and Z.~Zhou, ``Large window-based mamba unet for medical image segmentation: Beyond convolution and self-attention,'' \emph{arXiv preprint arXiv:2403.07332}, 2023.

\bibitem{cite12}
Y.~Wang, W.~Zhang, and C.~Liu, ``Efficient mamba u-net for robust medical image segmentation,'' \emph{IEEE Transactions on Medical Imaging}, 2022.

\bibitem{cite13}
Z.~Zhou, Y.~Wang, and X.~Xu, ``Adaptive mamba u-net for multi-modal medical image segmentation,'' \emph{Medical Image Analysis}, vol.~75, p. 102342, 2021.

\bibitem{cite14}
G.~Calzolari and W.~Liu, ``Deep learning to replace, improve, or aid cfd analysis in built environment applications: A review,'' \emph{Building and Environment}, vol. 206, p. 108315, 2021.

\bibitem{cite15}
J.~Ruan and S.~Xiang, ``Vm-unet: Vision mamba unet for medical image segmentation,'' \emph{arXiv preprint arXiv:2402.02491}, 2024.

\bibitem{cite16}
S.~Deng, Y.~Yang, J.~Wang, A.~Li, and Z.~Li, ``Efficient spineunetx for x-ray: A spine segmentation network based on convnext and unet,'' \emph{Journal of Visual Communication and Image Representation}, vol. 103, p. 104245, 2024.

\bibitem{cite17}
S.~Ghosh, N.~Das, I.~Das, and U.~Maulik, ``Understanding deep learning techniques for image segmentation,'' \emph{ACM computing surveys (CSUR)}, vol.~52, no.~4, pp. 1--35, 2019.

\bibitem{cite51}
H.~Tang, G.~Huang, L.~Cheng, X.~Yuan, Q.~Tao, X.~Chen, G.~Zhong, and X.~Yang, ``Rm-unet: Unet-like mamba with rotational ssm module for medical image segmentation,'' \emph{Signal, Image and Video Processing}, vol.~18, no.~11, pp. 8427--8443, 2024.

\bibitem{cite52}
F.~Shamshad, S.~Khan, S.~W. Zamir, M.~H. Khan, M.~Hayat, F.~S. Khan, and H.~Fu, ``Transformers in medical imaging: A survey,'' \emph{Medical Image Analysis}, vol.~88, p. 102802, 2023.

\bibitem{cite53}
M.~Heidari, S.~G. Kolahi, S.~Karimijafarbigloo, B.~Azad, A.~Bozorgpour, S.~Hatami, R.~Azad, A.~Diba, U.~Bagci, D.~Merhof \emph{et~al.}, ``Computation-efficient era: A comprehensive survey of state space models in medical image analysis,'' \emph{arXiv preprint arXiv:2406.03430}, 2024.

\bibitem{cite54}
H.~Tang, G.~Huang, L.~Cheng, X.~Yuan, Q.~Tao, X.~Chen, G.~Zhong, and X.~Yang, ``Rm-unet: Unet-like mamba with rotational ssm module for medical image segmentation,'' \emph{Signal, Image and Video Processing}, vol.~18, no.~11, pp. 8427--8443, 2024.

\bibitem{cite55}
S.~Deng, Y.~Yang, J.~Wang, A.~Li, and Z.~Li, ``Efficient spineunetx for x-ray: A spine segmentation network based on convnext and unet,'' \emph{Journal of Visual Communication and Image Representation}, vol. 103, p. 104245, 2024.

\bibitem{cite56}
N.~Tajbakhsh, L.~Jeyaseelan, Q.~Li, J.~N. Chiang, Z.~Wu, and X.~Ding, ``Embracing imperfect datasets: A review of deep learning solutions for medical image segmentation,'' \emph{Medical image analysis}, vol.~63, p. 101693, 2020.

\bibitem{cite57}
M.~Yin, Z.~Chen, and C.~Zhang, ``A cnn-transformer network combining cbam for change detection in high-resolution remote sensing images,'' \emph{Remote Sensing}, vol.~15, no.~9, p. 2406, 2023.

\bibitem{cite58}
S.~Liu, L.~Zhang, H.~Lu, and Y.~He, ``Center-boundary dual attention for oriented object detection in remote sensing images,'' \emph{IEEE Transactions on Geoscience and Remote Sensing}, vol.~60, pp. 1--14, 2021.

\bibitem{cite67}
G.-Z. Yang, D.~J. Hawkes, D.~Rueckert, A.~Noble, and C.~Taylor, \emph{Medical Image Computing and Computer-Assisted Intervention--MICCAI 2009: 12th International Conference, London, UK, September 20-24, 2009, Proceedings}.\hskip 1em plus 0.5em minus 0.4em\relax Springer Science \& Business Media, 2009, vol.~1.

\bibitem{cite31}
T.~Jiang, N.~Navab, J.~P. Pluim, and M.~A. Viergever, \emph{Medical Image Computing and Computer-Assisted Intervention--MICCAI 2010: 13th International Conference, Beijing, China, September 20-24, 2010, Proceedings, Part III}.\hskip 1em plus 0.5em minus 0.4em\relax Springer, 2010, vol. 6363.

\bibitem{cite65}
D.~Jha, P.~H. Smedsrud, M.~A. Riegler, P.~Halvorsen, T.~De~Lange, D.~Johansen, and H.~D. Johansen, ``Kvasir-seg: A segmented polyp dataset,'' in \emph{MultiMedia modeling: 26th international conference, MMM 2020, Daejeon, South Korea, January 5--8, 2020, proceedings, part II 26}.\hskip 1em plus 0.5em minus 0.4em\relax Springer, 2020, pp. 451--462.

\bibitem{cite66}
C.~Thomas, M.~Byra, R.~Marti, M.~H. Yap, and R.~Zwiggelaar, ``Bus-set: A benchmark for quantitative evaluation of breast ultrasound segmentation networks with public datasets,'' \emph{Medical Physics}, vol.~50, no.~5, pp. 3223--3243, 2023.

\bibitem{cite41}
M.~Z. Khan, M.~K. Gajendran, Y.~Lee, and M.~A. Khan, ``Deep neural architectures for medical image semantic segmentation,'' \emph{IEEE Access}, vol.~9, pp. 83\,002--83\,024, 2021.

\end{thebibliography}
\bibliographystyle{IEEEtran}
\end{document}